\documentstyle[12pt,epsfig,a4]{article} 
\def\bild#1#2{    
        \vspace*{-5mm}
        \begin{center}
        \begin{math}
        \epsfxsize#2cm
        \epsffile{#1}
        \end{math}
        \end{center}  }
\newcommand{\vs}{\vspace{-0.25cm}}
\begin{document} 

\begin{center}
\large{\bf Dominant $2\pi\gamma$-exchange nucleon-nucleon interaction: 
Spin-spin and tensor potentials}

\medskip

N. Kaiser\\

\smallskip

{\small Physik Department T39, Technische Universit\"{a}t M\"{u}nchen,
    D-85747 Garching, Germany}

\end{center}

\medskip

\begin{abstract}
We calculate at two-loop order in chiral perturbation theory the 
electromagnetic corrections to the two-pion exchange nucleon-nucleon
interaction that is generated by the isovector spin-flip $\pi\pi NN$
contact-vertex proportional to the large low-energy constant $c_4\simeq 3.4\,
$GeV$^{-1}$. We find that the respective $2\pi\gamma$-exchange potentials 
contain sizeable isospin-breaking components which reach up to $-4\%$ of 
corresponding isovector $2\pi$-exchange potentials. The 
typical values of these novel charge-independence breaking spin-spin and 
tensor potentials are $-0.11\,$MeV and $0.09\,$MeV, at a nucleon distance of 
$r=m_\pi^{-1}=1.4\,$fm. The charge-symmetry breaking spin-spin and tensor 
potentials come out a factor of $2.4$ smaller. Our analytical results for 
these presumably dominant isospin-violating spin-spin and tensor NN-forces are 
in a  form such that they can be easily implemented into phase-shift analyses 
and few-body  calculations.            
\end{abstract}


\medskip

To be published in: {\it The Physical Review C (2006), Brief Reports}

\bigskip
Isospin-violation in the nuclear force is a subject of current interest. 
Significant advances in the understanding of nuclear isospin-violation have 
been made in the past years by employing methods of effective field theory 
(in particular chiral perturbation theory). Van Kolck et al.\,\cite{kolck}
were the first to calculate (in a manifestly gauge-invariant way) the 
complete leading-order pion-photon exchange nucleon-nucleon interaction. In 
addition, the charge-independence and charge-symmetry breaking effects arising
from the pion mass difference $m_{\pi^+}- m_{\pi^0} = 4.59\,$MeV and the
nucleon mass difference $M_n- M_p = 1.29\,$MeV on the (leading order) two-pion
exchange NN-potential have been worked out in Refs.\,\cite{cib,csb}. Recently,
Epelbaum et al.\,\cite{evgenisobr} have continued this line of approach by
deriving the subleading isospin-breaking $2\pi$-exchange NN-potentials and
classifying the relevant isospin-breaking four-nucleon contact terms. 

In a recent work \cite{c3ppg} we have calculated the electromagnetic (i.e. 
one-photon exchange) corrections to the dominant two-pion exchange 
NN-interaction. The latter comes in  form of a strongly attractive 
isoscalar central potential generated by a one-loop triangle diagram involving 
the $\pi\pi NN$ contact-vertex proportional to the large low-energy constant 
$c_3 \simeq -3.3\,$GeV$^{-1}$. The dynamics behind this large value of $c_3$ is
the excitation of the low-lying $\Delta(1232)$-resonance. It has been found in
ref.\cite{c3ppg} that this particular class of two-loop $2\pi\gamma$-exchange
diagrams (proportional to $c_3$) leads to sizeable charge-independence and 
charge-symmetry breaking central NN-potentials which amount to $0.3\,$MeV at a
nucleon distance of $r = m_\pi^{-1} =1.4\,$fm. These are in fact the largest 
isospin-violating NN-potentials derived so far from long-range pion-exchange 
\cite{pigapot}. The effects from the other equally large low-energy constant 
$c_2 \simeq 3.3\,$GeV$^{-1}$ have also been studied in ref.\cite{c3ppg}. 
Isospin-violating potentials, $\widetilde V_C^{\rm(cib)}(m_\pi^{-1}) = -23.4\,
$keV and $V_C^{\rm(csb)}(m_\pi^{-1}) = -14.8\,$keV, more than an order of 
magnitude smaller have only been found. A reason for this suppression 
of the effects from the $c_2$-vertex is that the relevant two-loop spectral 
function stems in this case from a $2\pi\gamma$ three-body phase
space integral whose integrand vanishes on the phase space boundary.  

The remaining large low-energy constant $c_4 \simeq 3.4\,$GeV$^{-1}$ enters in
an isovector spin-flip $\pi\pi NN$ contact vertex. When inserted into the
two-loop $2\pi\gamma$-exchange diagrams the $c_4$-vertex generates 
isospin-violating spin-spin and tensor potentials. It is the purpose of the 
present short paper to calculate these remaining contributions to the
isospin-violating NN-potential along the same lines as in ref.\cite{c3ppg}
(i.e. by computing analytically the two-loop spectral functions). We do indeed 
find sizeable charge-independence breaking spin-spin and tensor potentials
with values of $-0.11\,$MeV and $0.09\,$MeV, at a nucleon distance of $r= 
m_\pi^{-1}=1.4\,$fm. These novel charge-independence breaking spin-spin and
tensor potentials are actually more than an order of magnitude larger than
their counterparts arising from (lowest order) pion-photon exchange (see
Table\,I in ref.\cite{pigapot}).

\bigskip
\medskip

\bild{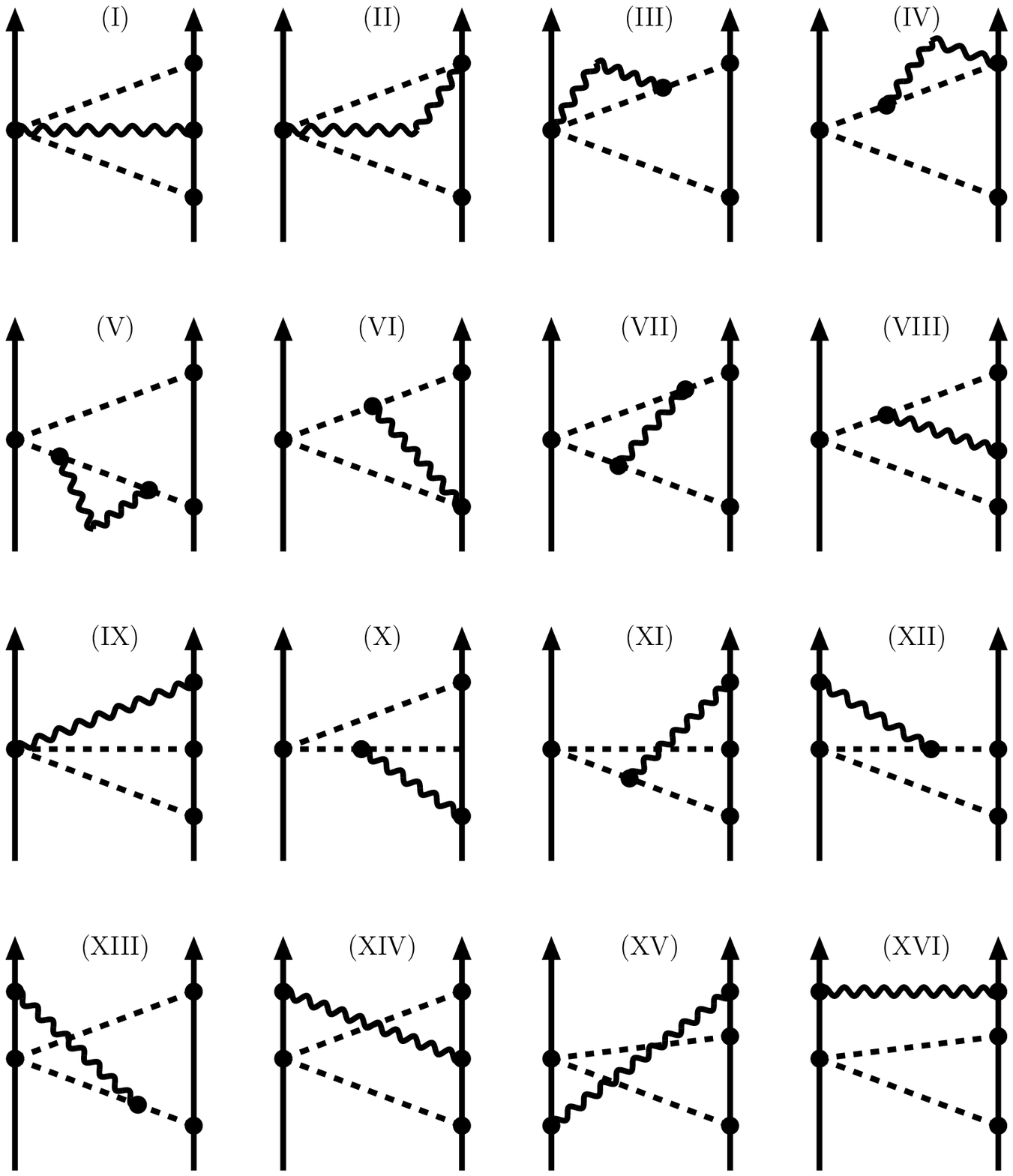}{14}
\vskip -0.2cm
{\it Fig.\,1: Electromagnetic corrections to the dominant isovector spin-spin
and tensor $2\pi$-exchange NN-interaction generated by the $\pi\pi NN$ 
contact-vertex $i c_4 f_\pi^{-2}\, \epsilon^{abc} \tau^c\, \vec \sigma\cdot
(\vec  q_a\times \vec q_b)$. Diagrams with the contact-vertex at the right
nucleon line and diagrams turned upside-down are not shown. The spectral 
functions Im$V_{S,T}(i\mu)$ are calculated by cutting the intermediate 
$\pi\pi\gamma$ three-particle state.}

\bigskip

In order to test their phenomenological
relevance these novel and exceptionally large isospin-violating spin-spin and
tensor potentials should be implemented into NN phase-shift analyses
\cite{rent1,rent2,rent3} and  few-body calculations.

Let us start with recalling the chiral $2\pi$-exchange NN-interaction. Its
dominant contribution to the isovector spin-spin and tensor channels comes
from a triangle diagram involving one $\pi\pi NN$ contact-vertex with 
momentum-dependent coupling: $i c_4 f_\pi^{-2}\,\epsilon^{abc} \tau^c \, \vec
\sigma \cdot (\vec q_a\times  \vec q_b)$. The corresponding potentials in 
coordinate space (obtained by a modified Laplace transformation) read
\cite{nnpot}: 
\begin{equation} \widetilde W_S^{(2\pi)}(r)={c_4 g_A^2 \over 48\pi^2 f_\pi^4}
\, {e^{-2z} \over r^6} \,(1+z)(3+3z+2z^2) \,, \end{equation}
\begin{equation} \widetilde W_T^{(2\pi)}(r)=-{c_4 g_A^2 \over 48\pi^2 f_\pi^4}
\, {e^{-2z} \over r^6} \,(1+z)(3+3z+z^2) \,, \end{equation}
where $z= m_\pi r$. The occurring parameters are: $g_A= 1.3$ (the nucleon 
axial-vector coupling constant), $f_\pi= 92.4\,$MeV (the pion decay constant),
$m_\pi=139.57\,$MeV (the charged pion mass) and the low-energy constant $c_4 = 
3.4\,$GeV$^{-1}$. The latter value has obtained in one-loop chiral
perturbation theory analyses of low-energy $\pi N$-scattering \cite{buett} and 
it has also been used (successfully) by Epelbaum et al.\,\cite{evgeni2} in 
calculations of NN-phase shifts at next-to-next-to-next-to-leading order in 
the chiral expansion. The numerical values in the first row of Tables I and II 
display the magnitude and $r$-dependence of these dominant isovector spin-spin 
and tensor $2\pi$-exchange potentials $\widetilde W_{S,T}^{(2\pi)}(r)$. It 
should also be noted that $\widetilde W_T^{(2\pi)}(r)$ reduces at intermediate 
distances, $1\,{\rm fm}\leq r\leq 2\,{\rm fm}$, the too strong isovector 
tensor potential from $1\pi$-exchange (see e.g. Fig.\,9 in ref.\cite{nnpot}).

Now we will add to the $2\pi$-exchange triangle diagram a photon line which
runs from one side to the other. There are five positions for the photon to
start on the left hand side and seven positions to arrive at the right hand
side. Leaving out those four diagrams which vanish in Feynman gauge (with
photon propagator proportional to $g_{\mu\nu}$) we get the 16 representative
diagrams shown in Fig.\,1. Except for diagram (I) these are to be understood  
as being duplicated by horizontally reflected partners. A further doubling of
the number of diagrams comes from interchanging the role of both nucleons. The 
two-loop diagrams in Fig.\,1 with the $c_4$-contact vertex included 
generate a contribution to the T-matrix of elastic NN-scattering of the form: 
\begin{equation}{\cal T}_{NN}=V_S(q)\,\,\vec \sigma_1 \cdot \vec \sigma_2 + 
V_T(q)\,\, \vec \sigma_1 \cdot \vec q \,\, \vec \sigma_2 \cdot \vec q \,, 
\end{equation}
with $\vec\sigma_{1,2}$ the usual spin-operators and $\vec q$ the momentum 
transfer between the two nucleons. We are interested here only in the
long-range parts of the associated coordinate-space potentials (disregarding
zero-range $\delta^3(\vec r\,)$-terms). For that purpose it is sufficient to 
know the spectral functions or imaginary parts of the two-loop diagrams. 
Making use of (perturbative) unitarity in the form of the Cutkosky 
cutting rule we can calculate the two-loop spectral functions as integrals of 
the (subthreshold) $\bar NN \to \pi\pi \gamma \to \bar NN $ transition 
amplitudes over the Lorentz-invariant $2\pi\gamma$ three-particle phase space.
In the (conveniently chosen) center-of-mass frame this leads to two angular
integrations and two integrals over the pion energies. Due to the heavy
nucleon limit ($M_N \to \infty$) and the masslessness of the photon 
($m_\gamma= 0$) several simplifications occur and therefore most of these 
integrations can actually be performed in closed analytical form. For a 
concise presentation of our results it is furthermore advantageous to consider 
two particular linear combinations of the spin-spin and tensor spectral
functions Im$V_{S,T}(i\mu)$ and to scale out all common (dimensionful) 
parameters: 
\begin{equation} 3\, {\rm Im} V_S(i\mu)- \mu^2 \, {\rm Im} V_T(i\mu) = {\alpha 
c_4 g_A^2 m_\pi^3 \over \pi (4f_\pi)^4}\,\,S_1\Big({\mu \over m_\pi} \Big) \,, 
\end{equation}
\begin{equation} {\rm Im} V_S(i\mu)- \mu^2\, {\rm Im} V_T(i\mu) = {\alpha 
c_4 g_A^2 m_\pi^3 \over \pi (4f_\pi)^4}\,\,S_2\Big({\mu \over m_\pi} \Big) \,. 
\end{equation}
Here, $\mu\geq 2m_\pi$ denotes the $\pi\pi\gamma$-invariant mass and $\alpha = 
1/137.036$ is the  fine-structure constant. Without going into further 
technical details, we enumerate now the contributions of the 16 representative 
$2\pi\gamma$-exchange diagrams shown in Fig.\,1 to the dimensionless spectral 
functions $S_1(u)$ and $S_2(u)$ which depend only on the dimensionless 
variable $u=\mu/m_\pi\geq 2$. We find three vanishing contributions: 
\begin{equation} S_1(u)^{(\rm I)} = S_1(u)^{(\rm VI)}= S_1(u)^{(\rm IX)} =0\,.
\end{equation}
In the Feynman gauge this property is obvious for diagrams (I) and (IX),
whereas one encounters for diagram (VI) a quite nontrivial double integral 
over the pion energies which surprisingly turns out to be equal to zero. The 
remaining nonvanishing contributions to the spectral function $S_1(u)$ read 
for $u\geq 2$:  
\begin{equation} S_1(u)^{(\rm II)} = \vec \tau_1 \cdot\vec\tau_2\,\bigg\{16-8u 
+{16\over u}  \ln(u-1) \bigg\}+\tau_1^3\tau_2^3 \,\bigg\{12u-2u^3 -{8\over u-1}
+{8\over u}  \ln(u-1) \bigg\}     \,, \end{equation} 
where $\tau_{1,2}^3$ denote the third components of the usual isospin
operators. 
\begin{equation} S_1(u)^{(\rm III)} = S_1(u)^{(\rm IV)} = (\vec \tau_1 \cdot  
\vec \tau_2+ \tau_1^3 \tau_2^3)\, \bigg\{{7u^3 \over 4} -3u^2 -{u\over 2} -1 + 
\bigg(8u-3u^3-{1\over u} \bigg)  \ln(u-1) \bigg\} \,, \end{equation} 
\begin{equation} S_1(u)^{(\rm V)} = (\vec \tau_1 \cdot \vec \tau_2+ \tau_1^3 
\tau_2^3)\,\bigg\{ 4u^2 - 3u^3 +6u-4 -{4\over u}(u^2-1)^2 \ln(u-1) \bigg\} \,, 
\end{equation} 
\begin{eqnarray} S_1(u)^{(\rm VII)}&=& \tau_1^3 \tau_2^3 \,\bigg\{{u^3\over 2}
+2u^2-11u +10+{2 \over u}(5u^4-14u^2+5) \ln(u-1)\nonumber \\ &&+4 \oint_1^{u/2}
\!\! dx \, {(2-ux)(u^2+2u x-4)\over (u-2x) y}\, \ln{u(x+y)-1 \over u(x-y)-1} 
\bigg\} \,, \end{eqnarray} 
with the abbreviation $y = \sqrt{x^2-1}$. 
\begin{eqnarray} S_1(u)^{(\rm VIII)}&=& (\tau_1^3+\tau_2^3-\tau_1^3 \tau_2^3 -
\vec \tau_1 \cdot \vec \tau_2) \, \bigg\{ {1\over 2u}(3+2u^2-u^4)\ln(u-1) 
\nonumber \\ && +{7u^3\over 8}-{5u^2\over 2}+{3u \over 4} +{3\over 2} +u 
\int_1^{ u/2}{dx\over y}(2-u x)\ln{u(x+y)-1\over u(x-y)-1}\bigg\}\,, 
\end{eqnarray} 
\begin{eqnarray} S_1(u)^{(\rm X)}&=&(\tau_1^3+\tau_2^3+\tau_1^3 \tau_2^3 +\vec 
\tau_1 \cdot \vec \tau_2)\,\bigg\{{1\over 2u}(u^4-4u^2-1)\ln(u-1) \nonumber \\ 
&&  + {u-2\over 8}(2+2u-u^2)+{1\over 4}(2-9u^2) \sqrt{u^2-4} +{2\over u}(u^4 
+1) \ln{ u + \sqrt{u^2-4} \over 2} \nonumber  \\ && + 2u \oint_1^{u/2} \!\! 
{dx \over u-2x}\bigg[2u y+(2-u^2) \ln{u-x+y \over u-x-y } \bigg] \bigg\}  \,, 
\end{eqnarray} 
\begin{eqnarray} S_1(u)^{(\rm XI)}&=& (\tau_1^3+\tau_2^3+3\tau_1^3 \tau_2^3 -
\vec \tau_1 \cdot \vec \tau_2) \, \bigg\{ 2u^2- {3u^3 \over 4} -{u\over 2} -1 
+\bigg( u -{1\over u} \bigg) \ln(u-1)\nonumber \\ && +{1\over 4}(2-9u^2) 
\sqrt{u^2-4} +{2\over u} (u^4+1) \ln{ u + \sqrt{u^2-4} \over 2}  + u
\int_1^{ u/2}{dx\over y}(u x-2) \nonumber \\ && \times \ln{u(x+y)-1\over u(x-y)
-1} + 2u \oint_1^{u/2} \!\! {dx \over u-2x}\bigg[2u y+(2-u^2) \ln{u-x+y \over
u-x-y } \bigg] \bigg\}  \,,  \end{eqnarray}  
\begin{eqnarray} S_1(u)^{(\rm XII)}&=&(\tau_1^3+\tau_2^3+\tau_1^3 \tau_2^3 +2
-\vec \tau_1 \cdot \vec \tau_2)\,\bigg\{{1\over 2u}(4u^2+1-u^4)\ln(u-1) 
\nonumber \\ &&  + {u-2\over 8}(u^2-2u-2)+{1\over 4}(2-9u^2) \sqrt{u^2-4}+{2 
\over u}(u^4 +1) \ln{ u + \sqrt{u^2-4} \over 2} \nonumber  \\ && + 2u 
\oint_1^{u/2} \!\! {dx \over u-2x}\bigg[2u y+(2-u^2) \ln{u-x+y \over u-x-y }
\bigg] \bigg\}  \,, \end{eqnarray}
\begin{eqnarray} S_1(u)^{(\rm XIII)}&=&(\tau_1^3+\tau_2^3- \tau_1^3 \tau_2^3 +
 2 +\vec \tau_1 \cdot \vec \tau_2)\,\bigg\{{1\over 2u}(u^4-4u^2-1)\ln(u-1) 
\nonumber \\ &&  + {u-2\over 8}(2+2u-u^2)+{1\over 4}(2-9u^2) \sqrt{u^2-4} 
+{2\over u}(u^4 +1) \ln{ u + \sqrt{u^2-4} \over 2} \nonumber  \\ && + 2u 
\oint_1^{u/2} \!\! {dx \over u-2x}\bigg[2u y+(2-u^2) \ln{u-x+y \over u-x-y }
\bigg] \bigg\}  \,, \end{eqnarray}  
\begin{equation} S_1(u)^{(\rm XIV)} = (\vec \tau_1 \cdot \vec \tau_2-1) \,  
\bigg\{ (u^2- 2) \sqrt{u^2-4} -{8\over u}\ln{ u +\sqrt{u^2-4} \over 2} 
\bigg\} \,, \end{equation}  
\begin{eqnarray} S_1(u)^{(\rm XV+ XVI)} &=& (\vec \tau_1 \cdot \vec \tau_2- 
\tau_1^3 \tau_2^3) \, \bigg\{ 4u-8 - {8\over u} \ln(u-1)\nonumber \\&&+(2-3u^2)
\sqrt{u^2-4} +{8\over u}(1+u^2)\ln{ u +\sqrt{u^2-4} \over 2}\bigg\} 
  \,. \end{eqnarray} 
The contribution from diagram (XV) together with the irreducible part of 
diagram (XVI) is proportional to the difference of their isospin factors on
the left nucleon line: $[1+\tau_1^3,\tau_1^c] \epsilon^{abc}= 2i(\tau_1^a
\delta^{b3}- \tau_1^b \delta^{a3}$). Furthermore, a closer inspection of the
spin-dependent factors reveals that the longitudinal spectral function
$S_2(u)$ receives only contributions from those diagrams where the photon does
not couple to a pion in flight. The few possible contributions to $S_2(u)$ 
from the diagrams (II), (XIV) and (XV+XVI) read for $u\geq 2$:
\begin{eqnarray} S_2(u)^{(\rm II)} &=& \vec \tau_1 \cdot\vec\tau_2\,\bigg\{
{u^3 \over 3}-{5u \over 2}+{10 \over 3}-{1\over u}-{2\over u^2}-{2\over u^3}
(u^2-1)^2 \ln(u-1) \bigg\} \nonumber \\ && +\tau_1^3\tau_2^3 \,\bigg\{{5u\over
2} -2 -{3\over u}-{6\over u^2}-{2\over u^3}(u^4+3)\ln(u-1) \bigg\}\,,
\end{eqnarray} 
\begin{equation} S_2(u)^{(\rm XIV)} = (1-\vec \tau_1 \cdot \vec \tau_2) \,  
\bigg\{ {u^2- 1 \over 6u^2}(u^2+6) \sqrt{u^2-4} +{2\over u^3}(2u^2-2-u^4) \ln{
u +\sqrt{u^2-4} \over 2}  \bigg\} \,, \end{equation}
\begin{eqnarray} S_2(u)^{(\rm XV+ XVI)} &=&  (\vec \tau_1 \cdot
\vec \tau_2- \tau_1^3 \tau_2^3) \, \bigg\{{2\over u^2}+{1\over u} -{10\over 3} 
+{5u \over 2}-{u^3 \over 3}+{2\over u^3} (u^2-1)^2 \ln(u-1) \nonumber \\ && + 
{u^2- 1 \over 6u^2}(u^2+6) \sqrt{u^2-4} +{2\over u^3}(2u^2-2-u^4) \ln{ u + 
\sqrt{u^2-4} \over 2} \bigg\} \,. \end{eqnarray}
We have also checked gauge-invariance. The (total) spectral functions $S_1(u)$
and $S_2(u)$ stay $\xi$-independent when adding a longitudinal part to the 
photon propagator: $g_{\mu\nu} \to g_{\mu\nu}+ \xi\, k_\mu k_\nu$. The 
''encircled'' integrals appearing in Eqs.(10,12--15) symbolize the following 
regularization prescription:  
\begin{equation} \oint_1^{u/2}\!\!dx\, {f(x)\over u-2x} =  \int_1^{u/2} \!\! 
dx\, { f(x)-f(u/2) \over u-2x} \,. \end{equation}
This regularization prescription eliminates from the contributions of some 
diagrams ((V), (VII), (X), (XI), (XII), (XIII)) to the spectral function  
$S_1(u)$ an infrared singularity arising from the emission of soft photons 
($\bar N N \to \pi\pi\gamma_{\rm soft}$). The singular factor $(u-2x)^{-1}$ 
originates from a pion propagator. The regularization prescription defined in 
Eq.(21) is equivalent to the familiar ''plus''-prescription employed commonly 
for parton splitting functions in order to eliminate there an analogous 
infrared singularity due to soft gluon radiation \cite{peskin}. It has also 
been used in our previous work in ref.\cite{c3ppg}. We note as an aside that
the non-elementary integrals ($\int_1^{u/2} dx \dots$) in Eqs.(10--15) can be
solved in terms of dilogarithms and squared logarithms.   

With the help of the spectral functions Im$V_{S,T}(i\mu)$ or $S_{1,2}(u)$ 
the $2\pi\gamma$-exchange spin-spin and tensor potentials in coordinate space  
$\widetilde V_{S,T}^{(2\pi\gamma)}(r)$ can be calculated easily via a modified
Laplace transformation:  
\begin{eqnarray} \widetilde V_S^{(2\pi\gamma)}(r) &=& {1\over 6\pi^2 r}
\int_{2m_\pi}^\infty \!d\mu\, \mu e^{-\mu r} \{\mu^2 \,{\rm Im} V_T(i\mu) -3\,
{\rm Im} V_S(i\mu) \}   \nonumber \\ &=& \widetilde V_S^{(0)}(r)+ \vec \tau_1 
\cdot \vec \tau_2 \, \widetilde W_S^{(0)}(r)+ \tau_1^3 \tau_2^3\, \widetilde 
V_S^{(\rm cib)}(r) +(\tau_1^3+ \tau_2^3)\, \widetilde V_S^{(\rm csb)}(r)\,,
\end{eqnarray} 
\begin{eqnarray} \widetilde V_T^{(2\pi\gamma)}(r) &=& {1\over 6\pi^2 r^3}
\int_{2m_\pi}^\infty \!d\mu\, \mu e^{-\mu r} \,(3+3\mu r+\mu^2 r^2)\, {\rm Im} 
V_T(i\mu)  \nonumber \\ &=& \widetilde V_T^{(0)}(r) + \vec \tau_1\cdot \vec 
\tau_2 \, \widetilde  W_T^{(0)}(r)+ \tau_1^3 \tau_2^3\, \widetilde V_T^{(\rm
cib)}(r)+(\tau_1^3+ \tau_2^3)\, \widetilde V_T^{(\rm csb)}(r)\,, \end{eqnarray}
where we have given in the second line of Eqs.(22,23) the decomposition into
isospin-conserving (0), charge-independence breaking (cib), and charge-symmetry
breaking (csb) parts. The numbers in the second to fifth row of Tables\,I and
II display the dropping of the spin-spin and tensor potentials $\widetilde 
V_{S,T}^{(0)}(r)$, $\widetilde  W_{S,T}^{(0)}(r)$, $\widetilde V_{S,T}^{(\rm 
cib)}(r)$ and $ \widetilde V_{S,T}^{(\rm csb)}(r)$ with the nucleon distance 
$r$ in the interval $1.0\,{\rm fm} \leq r \leq 1.9\,{\rm fm}$. One notices
quite sizeable charge-independence breaking (cib) spin-spin and tensor 
potentials. Their magnitudes (for example, $-0.11\,$MeV and $0.09$\,MeV at a 
nucleon distance of $r=m_\pi^{-1}=1.4\,$fm) are exceptionally large in 
comparison to all so far known isospin-breaking spin-spin and tensor
potentials generated by long-range pion-exchange. In particular, the 
potentials $\widetilde V_{S,T}^{(\rm  cib)}(r)$ exceed their counterparts from 
(leading and next-to-leading order) one-loop $\pi\gamma$-exchange by more than 
one order of magnitude (see herefore Table\,I in ref.\cite{pigapot}). As can 
be seen from the last row in Tables\,I and II the next largest $2\pi\gamma
$-exchange potentials are the charge-symmetry breaking (csb) ones. The
spin-spin and tensor potentials $ \widetilde V_{S,T}^{(\rm csb)}(r)$ reach
about $40\%$ of the charge-independence breaking ones and they have throughout
the same sign as $\widetilde V_{S,T}^{(\rm cib)}(r)$. We note also that in 
each case the numerically dominant contributions are coming from the diagrams
(V), (VII), (X)--(XIII). Chiral power counting \cite{evgenisobr} is obviously 
respected in our calculation since all contributions to the spectral functions 
Im$V_{S,T}(i\mu)$ scale in the same way with fine-structure constant $\alpha$
and the small mass and momentum scales $m_\pi$ and $\mu$. The sizable values
of the isospin-violating spin-spin and tensor potentials $\widetilde V_{S,T
}^{(\rm cib,csb)}(r)$ are due to the large value of the low-energy constant 
$c_4=3.4 \,$GeV$^{-1}$ and the large number of contributing diagrams.  

\begin{table}[hbt]
\begin{center}
\begin{tabular}{|c|cccccccccc|}
\hline 
$r$~[fm]& 1.0 & 1.1 & 1.2 & 1.3 & 1.4 & 1.5 & 1.6 & 1.7 & 1.8 & 1.9  
\\ \hline
$\widetilde W_S^{(2\pi)}$ [MeV]  & 24.95 & 13.61 & 7.78 & 4.63 &2.84 & 1.80 
   & 1.16 & 0.769 & 0.518 & 0.354 \\
$\widetilde V_S^{(0)}$ [keV] & --545 & --273 & --144 &--79.3 & --45.4 & --26.8
& --16.3 &  --10.1 & --6.44  & --4.17 \\
$\widetilde W_S^{(0)}$ [keV] & 240 & 126 & 68.9 & 39.5 & 23.4 & 14.3 & 8.97 &
5.76 & 3.77 & 2.51 \\
$\widetilde V_S^{(\rm cib)}$ [keV] & --1307 & --660 & --351 & --195 & --112 &
--66.8 & --40.9 & --25.6 & --16.4 & --10.7 \\ 
  $\widetilde V_S^{(\rm csb )}$ [keV] & --559 & --280 & --148 & --81.9& --47.0
& --27.8 & --16.9 & --10.6 & --6.72 & --4.37 \\ \hline
\end{tabular}
\end{center}
\vskip -0.2cm
{\it Table I: The dominant isovector  $2\pi$-exchange spin-spin potential 
$\widetilde W_S^{(2\pi)}(r)$, and electromagnetic corrections to it, as a 
function of the nucleon distance $r$. The values in the fourth and fifth row
correspond to the isospin-violating spin-spin potentials  $\widetilde V_S^{(\rm
cib)}(r)$ and $\widetilde V_S^{(\rm csb)}(r)$.}      
\end{table}

\begin{table}[hbt]
\begin{center}
\begin{tabular}{|c|cccccccccc|}
\hline 
$r$~[fm]& 1.0 & 1.1 & 1.2 & 1.3 & 1.4 & 1.5 & 1.6 & 1.7 & 1.8 & 1.9  \\ \hline
$\widetilde W_T^{(2\pi)}$ [MeV] &--22.91 &--12.35 & --6.98 &--4.10 & --2.49 & 
--1.56   & --0.996 &--0.652 &--0.434 & --0.294 \\
$\widetilde V_T^{(0)}$ [keV] & 452 & 224 & 117 & 63.9 & 36.2 & 21.2 
& 12.8 &  7.89 & 4.98 & 3.20  \\
$\widetilde W_T^{(0)}$ [keV] & --210 & --108 & --59.0 & --33.4 & --19.6 &
--11.9 & --7.38 &  --4.69 & --3.05 & --2.01 \\
$\widetilde V_T^{(\rm cib)}$ [keV] & 1098 & 549 & 289 & 159 & 91.0 & 53.7 &
32.5 & 20.2 & 12.8 & 8.31 \\ 
  $\widetilde V_T^{(\rm csb )}$ [keV]  & 469 & 233 & 122 & 66.7 & 37.9 & 22.3
  & 13.4 & 8.31 & 5.26 & 3.39 \\ \hline
\end{tabular}
\end{center}
\vskip -0.2cm
{\it Table II: The dominant isovector  $2\pi$-exchange tensor potential 
$\widetilde W_T^{(2\pi)}(r)$, and electromagnetic corrections to it, as a 
function of the nucleon distance $r$. The values in the fourth and fifth row
correspond to the isospin-violating tensor potentials  $\widetilde V_T^{(\rm
cib)}(r)$ and $\widetilde V_T^{(\rm csb)}(r)$.}      
\end{table}

In summary, we have calculated in this work the electromagnetic corrections to 
the two-pion exchange nucleon-nucleon interaction that is generated by the 
isovector spin-flip $\pi \pi NN$ contact-vertex proportional to the large 
low-energy constant $c_4=3.4\,$GeV$^{-1}$. 
Our analytical results for these novel and exceptionally large
isospin-violating spin-spin and tensor potentials  have been presented  in a 
form such that they can be easily implemented into future NN-phase shift
analyses and few-body calculations.

\end{document}